\documentclass[runningheads,a4paper]{llncs}

\usepackage{fullpage}
\usepackage{float}

\usepackage{amssymb}		
\usepackage{algorithm}
\usepackage{algcompatible}
\usepackage{tikz} \usetikzlibrary{chains,positioning,scopes,arrows,plotmarks} 
\usepackage{pgfplots}
\usepackage{tabto}		
\usepackage{rotating}		

\newcommand{\remove}[1]{}

\newcommand{\bac}{\backslash}
\newcommand{\seb} {\subseteq}
\newcommand{\matN}{\mathbb N}

\newcommand{\cL}{{\cal L}}
\newcommand{\cW}{{\cal W}}

\begin{document}

\mainmatter

\title{Power indices of influence games and new centrality measures\\for social networks\thanks{This work is partially supported by 2009SGR--1137 (ALBCOM).}}
\author{Xavier Molinero\inst{2}\thanks{Also partially funded by grant MTM2012--34426 of the ``Spanish Science and Innovation Ministry''.} \and Fabi\'an Riquelme\inst{1}\thanks{Also partially funded by grant BecasChile of the ``National Commission for Scientific and Technological Research of Chile'' (CONICYT) of the Chilean Government.} \and Maria Serna\inst{1}\thanks{Also partially funded by grant TIN2007--66523 (FORMALISM)  of ``Ministerio de Ciencia e Inovaci\'on y el Fondo Europeo de Desarrollo Regional''.}}

\institute{Departament de Llenguatges i Sistemes Inform\`atics, UPC, Barcelona, Spain.\\
	\and Department of Applied Mathematics III, UPC, Manresa, Spain.\\
	\email{xavier.molinero@upc.edu, farisori@lsi.upc.edu, mjserna@lsi.upc.edu}
}

\date{}

\maketitle

\begin{abstract}
In social network analysis, there is a common perception that influence is relevant to determine the global behavior of the society and thus it can be used to enforce cooperation by targeting an adequate initial set of individuals or to analyze global choice processes. Here we propose centrality measures that can be used to analyze the relevance of the actors in process related to spread of influence. In \cite{MRS12} it was considered a multiagent system in which the agents are eager to perform a collective task depending on the perception of the willingness to perform the task of other individuals. The setting is modeled using a notion of simple games called influence games. Those games are defined on graphs were the nodes are labeled by their influence threshold and the spread of influence between its nodes is used to determine whether a coalition is winning or not.

Influence games provide tools to measure the importance of the actors of a social network by means of classic power indices and provide a framework to consider new centrality criteria. In this paper we consider two of the most classical power indices, i.e., Banzhaf and Shapley-Shubik indices, as centrality measures for social networks in influence games. Although there is some work related to specific scenarios of game-theoretic networks, here we use such indices as centrality measures in any social network where the spread of influence phenomenon can be applied. Further, we define {\em new} centrality measures such as {\em satisfaction} and {\em effort} that, as far as we know, have not been considered so far.

Besides the definition we perform a comparison of the proposed measures with other three classic centrality measures, {\em degree}, {\em closeness} and {\em betweenness}. To perform the comparison we consider three social networks. We show that in some cases our measurements provide centrality hierarchies similar to those of other measures, while in other cases provide different hierarchies.
\keywords{Social Network, Centrality, Power index, Spread of Influence, Influence game, Simple game}
\end{abstract}

\section{Introduction}
\label{sec:intro}

Social network analysis is a multidisciplinary field related to sociology, computer science and mathematics, among other disciplines. In the last decades the field has grown extensively with the development of Internet and the emergence of online social networks.
One of the most studied concepts in social network analysis is {\em centrality}, that has to do with measuring how structurally important is an actor within a social network \cite{Fre79,BE06,ST11}. There are several centrality measures that provide different relevance criteria for the nodes within the network \cite{WF94,LM07}.
However, one of the major challenges for a successful implementation of network management activities, such as viral marketing, is the identification of key persons with a central structural position within the network. For this purpose, social network analysis provides a lot of measures for quantifying a member's interconnectedness within social networks, providing strongly differing results with respect to the quality of the different centrality measures. See \cite{LFH10} for a critical review in this context.

In this paper we consider four different centrality measures, focused directly on these kind of applications, where trends, ideas, fashions, etc. can be propagated through the network starting from a given set of actors. The first two measures correspond to classic power indices coming from the field of simple games and voting theory, namely the Banzhaf and Shapley-Shubik power indices, which can be bring to this centrality context through the use of {\em influence games} \cite{MRS12}, a way to represent simple games as if they were social networks.
There are some previous work where the Shapley-Shubik index is used as centrality measure for specific game-theoretic networks~\cite{SN10,MASRJ13}, but as far as we know, the Banzhaf index has not been used before for this proposal.

The other two measures are the {\em effort} and the {\em satisfaction}, {\em new} centrality measures that take advantage directly from the notion of influence games.
While effort's centrality measures the effort required to make the social network follow the opinion of an individual, satisfaction's centrality measures the level of satisfaction of each individual, so that it is influential if and only if it is taken into account.
Besides the necessary definitions, we perform an experimental comparison between these new centrality measures and three classic ones, which are {\em degree}, {\em closeness} and {\em betweenness}. We compare them on some simple real social networks for which a computation can be performed in reasonable time.\\

Influence games are strongly related with the notion of {\em spread of influence}, which describes the ways in which people influence each other through their interactions in a social network. This is certainly an intuitive and also well known phenomenon in social network analysis \cite{EK10} that is studied in other topics of this area, like the homophily phenomenon \cite{McPSC01,ST11}. It has also received a lot of attention in the last decade in the computer science community, a well as in other areas like viral marketing \cite{AM11,Jac08,EK10}.

Motivated by viral marketing and other applications, the problem that has been usually studied is the {\em influence maximization problem}, initially introduced by \cite{DR01,RD02} and further developed in \cite{KKT03,ES07}. This problem addresses the question of finding a set with at most $k$ players having maximum influence, and it is {\sc NP}-hard \cite{DR01}, unless additional restrictions are considered, in which case some generality of the problem is lost \cite{RD02}.
Two general models for spread of influence were defined in~\cite{KKT03}: the first one is the {\em linear threshold model}, based in the first ideas of \cite{Gra78,Sch78}, and the second one is the {\em independent cascade model}, created in the context of marketing by \cite{GLM01,GLM01b}.
Models for influence spread in the presence of multiple competing products have also been proposed and analyzed \cite{BKS07,BFO10,AM11}. In such a setting there is also work done towards analyzing the problem from the point of view of non-cooperative game theory. Non-cooperative influence games were defined in 2011 by Irfan and Ortiz~\cite{IO11}. Those games, however, analyze the strategic aspects of two firms competing on the social network and differ from our proposal.

Regarding power indices, in simple game theory a {\em power index} is a measure of the importance of the players of a game. They have been extensively studied in voting systems, sociology and economics.
Most studies are related to voting systems \cite{Azi09}, analyzing the computational complexity of calculating power indices in certain subclasses of simple games \cite{Kei08}, defining new power indices to represent the relevance of players under different considerations \cite{Fre12}, analogously to the works which define new centrality measures, like the present one.
As far as we know, this is the first approach to apply power indices as centrality measures for social networks. In fact, traditional centrality measures are not focused on social networks contemplating spread of influence processes.
However, the idea of using power indices as measures in networks is not new. The interested reader is referred to \cite{BR09} for studies concerning power indices in flow networks and to \cite{BRS11,BRS12} for measurements of power and satisfaction in societies. 

On the other hand, our {\em satisfaction} measure is inspired on the concept of {\em opinion leadership}, which has received considerable attention in sociology and marketing. It rose out of the two-step flow of communication theory introduced by \cite{KL55,LBG68}.
The existence ---or non-existence--- of opinion leaders in a society has a considerable impact on environments of marketing and politics.
These issues have been addressed recently on theoretical grounds by \cite{BRS11,BRS12}.
In these papers the authors consider a society formed by a social network with two layers, leaders and followers ---they also consider independent actors, who are not related to the other actors in the network---, and they introduce the measures of {\em satisfaction} and {\em power}.
The first one measures the ability of each actor ``to affect the state of the society concerning a specific outcome'', while the second tells us ``to which degree members of the society can be expected to end up with an outcome that they like''.
Interestingly, influence games can be saw as a generalization of this leader-follower model, so that a version of the satisfaction measure can be used as a centrality measure for this work. We do not consider here the measure of power, because it is very similar to that of satisfaction and provides similar results.\\

A social network can be represented by a graph where each node is an actor, individual, agent or player, and each edge connecting two nodes represents an interpersonal tie between the respective actors.
These graphs, depending on what we want to represent and on the complexity of the system, may be directed or undirected, and be labeled or not. For instance, in graphs with labeled edges, high weights on edges may mean strong interpersonal ties between the respective actors, and viceversa.
In some cases, the network can also be dynamic, changing its topology or properties over time.

In this paper, we consider static networks, defined beforehand, so that the number of nodes remains unchanged and there is no creation, deletion nor strengthening of interpersonal ties.
We also consider directed edges, so that they also represent the degree of influence of one actor over another one.

We consider three simple real social networks to perform a comparison on the proposed centrality measures with some traditional ones.
The first one, {\em monkeys' interaction}, corresponds to a unlabeled and undirected graph \cite{EB99,LM07}; the second one, {\em dining-table partners}, is an edge-labeled directed graph \cite{Mor60,dNMB04}; and the third one, {\em student Government discussion}, is an edge-labeled and vertex-labeled directed graph \cite{Hle93,dNMB04}. For the first two cases we take some additional considerations, while the last one corresponds exactly to an influence game. Our experimental results do not contradict the relevance criteria provided by traditional centrality measures like {\em degree centrality}, {\em closeness} or {\em betweenness}. In some cases such measurements are similar to our measurements, returning expected results for a reasonable measure of centrality.
However, there are also cases where the results have been quite different from traditional measures, which may also differ significantly, as is the case of the Student Government discussion network, or the Monkeys' interaction network for the effort centrality measure.
For these cases, new reasonable centrality measures provide new approaches and insights for social network analysis.
Moreover, the new centrality measures could be very eloquent for edge-labeled and vertex-labeled directed graphs, which is not supported by most measures of centrality \cite{Bor05}.

The paper is organized as follows: In Section~\ref{sec2} we recall concepts related to social networks and traditional centrality measures. In Section~\ref{sec3} we define the main concepts needed for our results, including our notion of {\em influence game}, which relates social networks with voting theory. In Section~\ref{sec4} we define some classic power indices, explaining why they can be consider as centrality measures, and we define a new centrality measure not based on power indices, the {\em effort} centrality measure. In Section~\ref{sec5} we compare all these new centrality measures with others on real and known cases of study. We finish with some conclusions and future work.

\section{Preliminars}
\label{sec2}

A {\em social network} is an edge-labeled graph $(G,w)$, where $G=(V,E)$ is a graph without loops, $V$ is the set of nodes representing individual, actors, players, etc., $E$ is the set of edges representing interpersonal ties between actors, and $w:E\to\matN$ is a {\em weight function} which assigns a weight to every edge, representing the strength of each interpersonal tie. An actor $i\in V$ has {\em influence} over another $j\in V$ iff $(i,j)\in E$.

The graph $G$ could be directed or undirected, and can have weighted nodes or not. Undirected graphs can be treated as symmetric directed graphs, considering that an undirected edge $\{i,j\}$ is the same that two directed edges $(i,j)$ and $(j,i)$.\\

The {\em centrality} of a node refers to its relative importance inside of a network, and depends of structural aspects at a global level. Centrality is one of the most studied concepts in network analysis, and since the late 1970s in social network analysis \cite{Fre79,FBW91}. There are several centrality measures \cite{KLPRTZ05} that provide different importance criteria to the nodes. Let $i\in V$ be an actor, three of the most well-known and widely applied are {\em degree centrality}, {\em closeness centrality} and {\em betweenness centrality}~\cite{WF94}, which in a normalized version are defined as follows~\cite{LM07}:

\begin{itemize}
  \item Degree centrality ($C_D$): It is based on the indegree or outdegree of each actor, i.e.,
        $$C_D^-(i)=\frac{deg^-(i)}{n-1} \mbox{\,\, or \,\,} C_D^+(i)=\frac{deg^+(i)}{n-1}.$$
	where $deg^-(i)= |\{j\in{}V\,|\,(j,i)\in{}E\}|$ is the number of vertices that goes to $i$, and $deg^+(i)= |\{j\in{}V\,|\,(i,j)\in{}E\}|$ is the number of vertices that goes from $i$.
	For undirected networks, $deg(i)=deg^-(i)=deg^+(i)$, so $C_D$ is without distinction $C_D^-$ and $C_D^+$.
  \item Closeness centrality ($C_C$): It is based on the inverse of the sum of shortest paths from $i$ to the other actors, i.e., let $D$ be the usual distance matrix of the network,
        $$C_C(i)=\frac{n-1}{\sum_{i\neq j}(D)_{ij}}.$$
	If there is no path from $i$ to $j$, we assume that $(D)_{ij}=n$.
  \item Betweenness centrality ($C_B$): Let $b_{jk}$ the number of shortest paths from the node $j$ until $k$, and $b_{jik}$ the number of these shortest paths that pass through $i$, then
        $$C_B(i)=\frac{1}{(n-1)(n-2)}\sum_{j\neq k}\frac{b_{jik}}{b_{jk}}.$$
	If there is no path from $j$ to $k$, we assume $\frac{b_{jik}}{b_{jk}}=0$.
\end{itemize}

There are several centrality measures based on the previous ones, such as the Katz centrality, Bonacich centrality, Hubbell centrality, Newman betweenness, between others \cite{ST11}. The differences between these variations are few, and do not involve a change of paradigm. Additionaly, there are other measures based on other ideas, like Eigenvector and Alpha centrality \cite{ST11}. Moreover, some of these measures were initially defined only for undirected graphs, but there also exist generalizations which consider weighted edges \cite{SZ89,OAS10}. However, most of them do not consider graphs with weighted nodes, which is necessary for our notion of spread of influence phenomenon.

\section{Social networks as simple games}
\label{sec3}

The following definition is based on the {\em linear threshold model} \cite{Gra78,Sch78,KKT03}. We follow notation from \cite{MRS12}.

\begin{definition}
An {\em influence graph} is a tuple $(G,w,f)$ where $(G,w)$ is a social network and $f:V\to\matN$ is a labeling function that quantifies how influenciable each actor is.
\end{definition}

Given an influence graph $(G,w,f)$ and an initial activation set $X\seb V$, the {\em spread of influence} is denoted by $F(X)$, where $F(X)\seb V$ is formed by the actors activated through an iterative process in which initially only the nodes in $X$ are activated. Let be $F^t(X)$ the set of nodes activated at some iteration $t$, then at the next $t+1$ iteration a node $i\in V$ will be activated iff:
$$\sum_{j\in F^t(X)} w((j,i)) \geq f(i).$$
In other words, a node $i$ is activated when the weights' sum of the activated nodes connected to this node $i$ is greater or equals to its capacity of influence. The process stops when no additional activation occurs, so that $F(X)=F^1(X)\cup\ldots\cup F^k(X)$, where $k\leq n$.

\begin{example}\label{ex1}
Figure \ref{fig1} shows the spread of influence $F(X)$ in an influence graph from the initial activation $X=\{a\}$.
In the first step we obtain $F^1(X)=\{a,c\}$, and in the second step ---the last one---, $F^2(X)=\{a,c,d\}$.

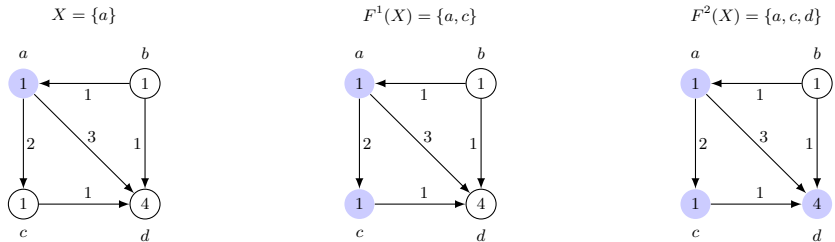
\begin{figure}[H]
\centering
\begin{minipage}[b]{0.23\linewidth}
\begin{center}
\begin{tikzpicture}[every node/.style={circle,scale=0.7}, >=latex]
\node[scale=1]		at (0.8,2.5){$X=\{a\}$};
\node[fill=blue!20](a)  at (0.0,1.6)[label=above:$a$] {1};
\node[draw](b) 		at (1.6,1.6)[label=above:$b$] {1};
\node[draw](c) 		at (0.0,0.0)[label=below:$c$] {1};
\node[draw](d)		at (1.6,0.0)[label=below:$d$] {4};
\draw[->] (a) to node {}(c);
\draw[->] (a) to node {}(d);
\draw[->] (b) to node {}(a);
\draw[->] (b) to node {}(d);
\draw[->] (c) to node {}(d);
\node[scale=1]	at (0.85,1.45){$1$};
\node[scale=1]	at (0.1,0.8){$2$};
\node[scale=1]	at (0.9,0.9){$3$};
\node[scale=1]	at (1.5,0.8){$1$};
\node[scale=1]	at (0.85,0.15){$1$};
\end{tikzpicture}
\end{center}
\end{minipage}
\qquad
\begin{minipage}[b]{0.23\linewidth}
\begin{center}
\begin{tikzpicture}[every node/.style={circle,scale=0.7}, >=latex]
\node[scale=1]		at (0.8,2.5){$F^1(X)=\{a,c\}$};
\node[fill=blue!20](a)  at (0.0,1.6)[label=above:$a$] {1};
\node[draw](b) 		at (1.6,1.6)[label=above:$b$] {1};
\node[fill=blue!20](c) 	at (0.0,0.0)[label=below:$c$] {1};
\node[draw](d)		at (1.6,0.0)[label=below:$d$] {4};
\draw[->] (a) to node {}(c);
\draw[->] (a) to node {}(d);
\draw[->] (b) to node {}(a);
\draw[->] (b) to node {}(d);
\draw[->] (c) to node {}(d);
\node[scale=1]	at (0.85,1.45){$1$};
\node[scale=1]	at (0.1,0.8){$2$};
\node[scale=1]	at (0.9,0.9){$3$};
\node[scale=1]	at (1.5,0.8){$1$};
\node[scale=1]	at (0.85,0.15){$1$};
\end{tikzpicture}
\end{center}
\end{minipage}
\qquad
\begin{minipage}[b]{0.23\linewidth}
\begin{center}
\begin{tikzpicture}[every node/.style={circle,scale=0.7}, >=latex]
\node[scale=1]		at (0.8,2.5){$F^2(X)=\{a,c,d\}$};
\node[fill=blue!20](a)  at (0.0,1.6)[label=above:$a$] {1};
\node[draw](b) 		at (1.6,1.6)[label=above:$b$] {1};
\node[fill=blue!20](c) 	at (0.0,0.0)[label=below:$c$] {1};
\node[fill=blue!20](d)	at (1.6,0.0)[label=below:$d$] {4};
\draw[->] (a) to node {}(c);
\draw[->] (a) to node {}(d);
\draw[->] (b) to node {}(a);
\draw[->] (b) to node {}(d);
\draw[->] (c) to node {}(d);
\node[scale=1]	at (0.85,1.45){$1$};
\node[scale=1]	at (0.1,0.8){$2$};
\node[scale=1]	at (0.9,0.9){$3$};
\node[scale=1]	at (1.5,0.8){$1$};
\node[scale=1]	at (0.85,0.15){$1$};
\end{tikzpicture}
\end{center}
\end{minipage}
\caption{Spread of influence ---colored nodes--- from the initial activation $X=\{a\}$.\label{fig1}}
\end{figure}
\end{example}

Now we define simple games, a useful mathematical structure used in voting theory, cooperative game theory, and many other topics like hypergraphs or monotone Boolean functions \cite{TZ99,Azi09,RP11}.

\begin{definition}
A {\em simple game} is a tuple $(N,\cW)$ where $N$ is a finite set of players and $\cW$ is a monotonic family of subsets of $N$ formed by the {\em winning coalitions}, such that if $X\in\cW$ and $X\seb Z$, then $Z\in\cW$.
\end{definition}

Based on simple games, we use a simplified definition of the given by \cite{MRS12}, considering that all the nodes can be initially activated.

\begin{definition}\label{def:inf-game}
An {\em influence game} is a tuple $(G,w,f,q)$ where $(G,w,f)$ is an influence graph and $q$ is a {\em quota} $0\leq q\leq |V|+1$. $X\seb V$ is a {\em winning coalition} iff $|F(X)|\geq q$, otherwise $X$ is a {\em losing coalition}.
\end{definition}

Since now, we assume that $N=V$ and $n=|N|$. Furthermore, we assume that for each isolated node $i$, $f(i)$ tends to infinity, which means that it can not be convinced in any way.
Note that influence games are also monotonic, because for any $X\seb V$, $i\in V$, if $|F(X)|\geq q$ then $|F(X\cup\{i\})|\geq q$, and if $|F(X)|<q$ then $|F(X\bac\{i\})|<q$. Thus, it is clear that every influence game is a simple game. The opposite is also true as it was shown by \cite{MRS12}.

\section{Power indices and new centrality measures}
\label{sec4}
A {\em power index} is a measure of the importance of the players of a game.
Power indices have been extensively studied in voting systems, sociology and economics.
Most studies are related to voting systems \cite{Azi09}, analyzing the computational complexity of calculating power indices in certain subclasses of simple games \cite{Kei08}, defining new power indices to represent the relevance of players under different considerations \cite{Fre12}, analogously to the works which define new centrality measures, like the present one.

The main power indices are the {\em Banzhaf index} \cite{Ban65} ---also called {\em Penrose-Banzhaf index} \cite{Pen46}--- and the {\em Shapley-Shubik index} \cite{SS54}.
Let $(N,\cW)$ be a simple game, and let $C_i=\{S\in\cW; S\bac\{i\}\notin\cW\}$ be the set of {\em blocking coalitions} for each $i\in N$.
A player is {\em critic} in a coalition if that coalition is a blocking.
The {\em Banzhaf value} of $i$ is $$\eta(i)=|C_i|$$ and the
{\em Shapley-Shubik value} of $i$ is $$\kappa(i)=\sum_{S\in C_i}(|S|-1)!\,(n-|S|)!$$
The normalized versions give the {\em Banzhaf index} ({\tt Bz}) and the {\em Shapley-Shubik index} ({\tt SS}), respectively,
$$\mbox{{\tt Bz}}(i)=\frac{\eta(i)}{\sum_{i\in N}\eta(i)} \mbox{ ~~ and ~~ {\tt SS}}(i)=\frac{\kappa(i)}{n!}.$$ 

Note that both power indices (seen as measures), like $C_B$, correspond to {\em medial measures} in the sense that they take as reference the sets of actors which pass through a given node instead of a given node which starts or ends some paths through the network, like it succeed with the {\em radial measures} as $C_D$ or $C_C$ \cite{ST11}.

Both power indices can be considered as centrality measures because an actor is more central in the network while more necessary is for generating of winning coalitions.

\begin{example}\label{ex:power}
Lets consider the influence game given by the influence graph of Figure \ref{fig1} and a quota $q=n$. Player $b$ is the unique critic player (i.e., $b$ belongs to any winning coalition), so $\eta(b)=8$ and {\tt Bz}$(b)=1$, while 
$\eta(j)=0$ and {\tt Bz}$(j)=0$ for $j\in\{a,c,d\}$. Further, as $|\{b\}|=1$, $|\{b,a\}| = |\{b,c\}| = |\{b,d\}|=2$, $|\{b,a,c\}| = |\{b,a,d\}| = |\{b,c,d\}|=3$ and $|\{b,a,c,d\}|=4$, then $\kappa(b)=24$ and {\tt SS}$(b)=1$, while {\tt SS}$(j)=0$ for $j\in\{a,c,d\}$.
\end{example}

The interested reader can see \cite{Fre11} and \cite{Kei08} to consider other power indices, i.e., the {\em Deegan-Packel index} \cite{DP78}, the {\em Holler index} \cite{Hol82}, the {\em Coleman indices} \cite{Col71} or the {\em Johnston index} \cite{Joh78} among many others.\\

On the other hand, influence games can also provide other {\em new} criteria to determine measures of centrality. For instance,
to consider the weights of the nodes of the smaller coalition in which an individual can be contained:
Let be $w(S)=\sum_{i\in S}f(i)$ the sum of the weights of the nodes of a given coalition $S\seb N$,
we define the minimum effort required by the system to choose the initial activation that contains a required individual, in such a way that this activation is a winning coalition:
$$\mbox{{\tt Effort}}(i)=\min\{w(S); S\cup\{i\}\in\cW\}.$$
When we apply such concept to influence games setting a quota $q$ then the {\em effort} is:
$$\mbox{{\tt Effort}}(i)=\min\{w(S); F(S\cup\{i\})\geq q\}.$$

Note that while greater is the required effort for a node, this node should be less central.
Therefore, we define the normalized version of the {\em effort centrality measure} ($C_E$) as follows:
$$C_E(i)=\frac{w(N)-\mbox{{\tt Effort}}(i)}{w(N)}.$$

\begin{example}
For the same influence game of Example~\ref{ex:power}, {\tt Effort}$(b)=1$, {\tt Effort}$(a)$ = {\tt Effort}$(c)=2$ and {\tt Effort}$(d)=5$, so $C_E(b)=6/7 > C_E(a)=C_E(c)=5/7 > C_E(d)=2/7$.
\end{example}

Moreover, \cite{BRS11,BRS12} define a {\em satisfaction score} to measure the ability to affect the state of an opinion leader-follower collective choice situation. Every instance of this model can be saw as an influence game formed by isolated nodes and a bipartite graph with two levels, such that there are only directed edges pointing from one level to the other. Without going into detail on this model, just note that influence games are a generalization of this. Thus, we define the {\em satisfaction centrality measure} ($C_S$), based on the satisfaction score, as follows.
Let $(G,w,f,q)$ be an influence game, $\cW_i=\{X\seb V(G); i\in X, F(X)\geq q\}$ and $\cL_{-i}=\{X\seb V(G); i\notin X, F(X)<q\}$, then:
$$C_S(i)=\frac{|\cW_i|+|\cL_{-i}|}{2^n}.$$

It is interesting to note that this satisfaction score, and therefore the new satisfaction measure, leaving aside losing coalitions would be the same that the well known solution concepts called {\em Chow parameters} \cite{Cho61,DS79}, deeply related with the {\em Holler index} \cite{Hol82}.

Note that we could define other measures based on $C_E$ or $C_S$, but they probably provide similar or less interesting results.
For instance, we could also define the effort without weights, based on the width parameter~\cite{Azi09}:
$$\mbox{{\tt Width}}(i)=\min\{|S|; F(S\cup\{i\})\geq q\}.$$
and then to consider
$$C_W(i)=\frac{n-\mbox{{\tt Width}}(i)}{n}.$$
However, this measure does not provide very interesting results, so it is not considered in this work.

\section{Cases of study}
\label{sec5}

We consider three simple real social networks to compare the new centrality measures {\tt Bz}, {\tt SS}, $C_E$ and $C_S$, with some traditional ones, $C_D$, $C_C$ and $C_B$.
The first one, {\em monkeys' interaction}, corresponds to a unlabeled and undirected graph; the second one, {\em dining-table partners}, is an edge-labeled directed graph; and the third one, {\em student Government discussion}, is an edge-labeled and vertex-labeled directed graph. For the first two cases we will take some additional considerations, while the last one corresponds exactly to an influence game as Definition~\ref{def:inf-game}.

\subsection{Monkeys' interaction}

\begin{figure}[ht]
\begin{center}
\begin{tikzpicture}[every node/.style={circle,scale=0.7}, >=latex]
\node[draw](1) at (1.2,5.9)[label=above:$1$] {};
\node[draw](3) at (1.7,3.5)[label=above:$3$] {};
\node[draw](4) at (5.0,5.3)[label=above:$4$] {};
\node[draw](5) at (0.1,4.2)[label=left :$5$] {};
\node[draw](7) at (3.0,6.0)[label=above:$7$] {};
\node[draw](8) at (0.3,5.2)[label=above:$8$] {};
\node[draw](9) at (0.0,3.1)[label=left :$9$] {};
\node[draw](10)at (0.2,2.0)[label=left :$10$]{};
\node[draw](11)at (2.5,4.3)[label=above:$11$]{};
\node[draw](12)at (3.6,4.0)[label=right:$12$]{};
\node[draw](13)at (5.2,3.0)[label=right:$13$]{};
\node[draw](14)at (1.0,1.2)[label=below:$14$]{};
\node[draw](15)at (3.2,1.0)[label=below:$15$]{};
\node[draw](17)at (4.5,1.8)[label=below:$17$]{};
\node[draw](2) at (0.0,0.0)[label=below:$2$] {};
\node[draw](6) at (1.0,0.0)[label=below:$6$] {};
\node[draw](16)at (2.0,0.0)[label=below:$16$]{};
\node[draw](18)at (3.0,0.0)[label=below:$18$]{};
\node[draw](19)at (4.0,0.0)[label=below:$19$]{};
\node[draw](20)at (5.0,0.0)[label=below:$20$]{};
\draw[-] (1) to node {}(3);
\draw[-] (1) to node {}(7);
\draw[-] (1) to node {}(8);
\draw[-] (1) to node {}(12);
\draw[-] (3) to node {}(4);
\draw[-] (3) to node {}(5);
\draw[-] (3) to node {}(7);
\draw[-] (3) to node {}(8);
\draw[-] (3) to node {}(9);
\draw[-] (3) to node {}(10);
\draw[-] (3) to node {}(11);
\draw[-] (3) to node {}(12);
\draw[-] (3) to node {}(13);
\draw[-] (3) to node {}(14);
\draw[-] (3) to node {}(15);
\draw[-] (3) to node {}(17);
\draw[-] (4) to node {}(13);
\draw[-] (4) to node {}(15);
\draw[-] (5) to node {}(8);
\draw[-] (7) to node {}(12);
\draw[-] (10)to node {}(12);
\draw[-] (10)to node {}(15);
\draw[-] (11)to node {}(12);
\draw[-] (12)to node {}(13);
\draw[-] (12)to node {}(14);
\draw[-] (12)to node {}(15);
\draw[-] (12)to node {}(17);
\draw[-] (13)to node {}(14);
\draw[-] (13)to node {}(15);
\draw[-] (13)to node {}(17);
\draw[-] (14)to node {}(15);
\end{tikzpicture}
\end{center}
\caption{Social network of monkeys recorded by \cite{EB99} and also used by \cite{LM07}.\label{fig2}}
\end{figure}
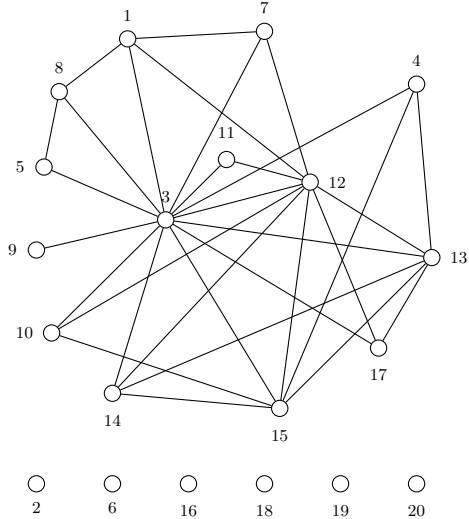

In \cite{EB99} is provided a network which represents the real interactions amongst a group of 20 monkeys observed during three months next to a river.
It corresponds to an undirected graph where an edge $\{i,j\}$ exists when monkeys $i$ and $j$ were witnessed together in the river. The graph is formed by 6 isolated nodes and a connected component of 14 nodes, as showed Figure \ref{fig2}.
The authors considered the centrality measures $C_D$, $C_C$ and $C_B$, as well as generalized versions of these measures for groups instead of individuals. Years later, \cite{LM07} use the same network to compare the previous results with the measure called {\em information centrality}.

In order to analize this network $((V,E),w)$ we assume that every undirected edge $\{i,j\}$ with $i,j\in V$ represents in fact two arcs $(i,j)$ and $(j,i)$ of $E$, so the graph is symmetric.
Moreover, we assume that the weight function is defined by $w(e)=1$, for all $e\in E$.
In the context of our work, this means that a monkey can influence and be influenced by other monkey iff they have interacted before.

To define from here an influence game we use the quota $q=14$, which corresponds to the maximum spread of influence which can be obtained from a monkey. This helps to obtain lower measures in isolated nodes, as it is to be expected from a centrality measure.

Now we consider the following {\em natural} labeling functions for every node $i\in V$:

\begin{itemize}
 \item Case 1 (C1): Minimum influence required to convincement, $f(i)=1$.
 \item Case 2 (C2): Average influence required to convincement, $f(i)=\lceil deg(i)/2\rceil$.
 \item Case 3 (C3): Majority influence required to convincement, $f(i)=\lfloor deg(i)/2\rfloor+1$.
 \item Case 4 (C4): Maximum influence required to convincement, $f(i)=deg(i)$.
\end{itemize}

The {\tt Bz}, {\tt SS}, $C_E$ and $C_S$ measures have been computed for all these cases: See Table \ref{tab:tab1}.
Note that all columns of the tables in this paper use a sufficient number of significant digits to distinguish those values which are ​​really different, from the ones that are equal.
Hence, only isolated nodes for {\tt Bz}, {\tt SS} and $C_E$, as well as the last column of $C_E$ assume a score exactly equal to zero.

Note that if we consider minimum influence required to convincement ---columns {\tt Bz}-C1, {\tt SS}-C1, $C_E$-C1 and $C_S$-C1---, then the new measures are not a good representatives, because as the spread of influence is fluid, i.e. actors does not require too many restrictions to form winning coalitions, then all the non-isolated nodes have the same value. However, for the other cases, when differences between influences are relevant, only the pair of monkeys $(10,17)$ and $(13,15)$ assume the same value for {\tt Bz}, {\tt SS} and $C_S$, allowing a more relevant hierarchization than for the other measures.

Leaving aside the case 1 for the new measures, there are other similarities between traditional measures and new ones.
In the same way than traditional measures, the most central monkey for {\tt Bz}, {\tt SS} and $C_S$ is~3.
The second score is for monkey 12, except for {\tt Bz}-C3 and {\tt SS}-C3, where it is replaced by monkeys 13 and 15.
Third score is for monkeys 13 and 15, except for {\tt Bz}-C3 and {\tt SS}-C3, where they are replaced, respectively, by monkey~14 and monkey~12,
and for $C_S$-C4, where they are replaced by monkey~8.
Finally, as expected, for all cases the less central non-isolated monkey is monkey~9, except for $C_S$-C2, in which case is monkey~5.
Similarities between some traditional measures and some of the new ones are shown in Figure \ref{fig:chart_monkeys}. This also holds for $C_S$, although the scores are always greater.

On the other hand, $C_E$ provides a different criterion of centrality, for instance, now monkey~3 is the less central for $C_E$-C2.

\begin{table*}[ht]
\begin{center}
\scalebox{0.7}{
\begin{tabular}{|c|c|c|c|c|c|c|c|c|c|c|c|c|c|c|c|c|c|c|c|}\cline{5-20}
\multicolumn{4}{c|}{\,} & \multicolumn{4}{c|}{\tt Bz} & \multicolumn{4}{c|}{\tt SS} & \multicolumn{4}{c|}{$C_E$} & \multicolumn{4}{c|}{$C_S$}\\\hline
Node &$C_D$&$C_C$ &$C_B$ &C1   &C2    &C3     &C4     &C1   &C2    &C3    &C4    &C1  &C2   &C3   &C4 &C1 &C2 &C3 &C4\\\hline
1    &0.21 &0.134 &0.006 &0.07 &0.038 &0.0708 &0.0885 &0.07 &0.025 &0.068 &0.075 &0.9 &0.43 &0.14 &0 &0.501 & 0.521 & 0.575 &0.598\\
2    &0.00 &0.050 &0.000 &0.00 &0.000 &0.0000 &0.0000 &0.00 &0.000 &0.000 &0.000 &0.0 &0.00 &0.00 &0 &0.500 &0.500 &0.500 &0.500 \\
3    &{\bf 0.68}&{\bf 0.143}&{\bf 0.260}&0.07&{\bf 0.156}&{\bf 0.1214}&{\bf 0.1730}&0.07&{\bf 0.219}&{\bf 0.150}&{\bf 0.192}&0.9&0.36 &0.07 &0 &0.501 & {\bf 0.589} &{\bf 0.644} &{\bf 0.736}\\
4    &0.16 &0.133 &0.000 &0.07 &0.059 &0.0673 &0.0343 &0.07 &0.047 &0.062 &0.044 &0.9 &{\bf 0.50} &0.14 &0 &0.501 & 0.537 &0.547 &0.580\\
5    &0.11 &0.132 &0.000 &0.07 &0.019 &0.0373 &0.0438 &0.07 &0.013 &0.032 &0.036 &0.9 &0.43 &0.14 &0 &0.501 & 0.510 &0.543 &0.578\\
6    &0.00 &0.050 &0.000 &0.00 &0.000 &0.0000 &0.0000 &0.00 &0.000 &0.000 &0.000 &0.0 &0.00 &0.00 &0 &0.500 &0.500 &0.500 &0.500 \\
7    &0.16 &0.133 &0.000 &0.07 &0.049 &0.0497 &0.0460 &0.07 &0.032 &0.043 &0.045 &0.9 &0.43 &0.14 &0 &0.501 & 0.528 &0.551 &0.583\\
8    &0.16 &0.133 &0.003 &0.07 &0.048 &0.0282 &0.0863 &0.07 &0.040 &0.024 &0.066 &0.9 &0.43 &0.07 &0 &0.501 & 0.527 &0.532 &{\bf 0.601}\\
9    &0.05 &0.131 &0.000 &0.07 &0.028 &0.0281 &0.0003 &0.07 &0.017 &0.022 &0.005 &0.9 &0.43 &0.14 &0 &0.501 & 0.516 &0.531 &0.548\\
10   &0.16 &0.133 &0.000 &0.07 &0.074 &0.0538 &0.0205 &0.07 &0.069 &0.050 &0.035 &0.9 &{\bf 0.50} &0.14 &0 &0.501 & 0.536 &0.555 &0.582\\
11   &0.11 &0.132 &0.000 &0.07 &0.037 &0.0470 &0.0035 &0.07 &0.023 &0.040 &0.016 &0.9 &{\bf 0.50} &0.14 &0 &0.501 & 0.520 &0.553 &0.574\\
12   &{\bf 0.47}&{\bf 0.139}&{\bf 0.060}&0.07&{\bf 0.154}&0.1004&{\bf 0.1671} &0.07 &{\bf 0.180}&{\bf 0.107}&{\bf 0.160} &0.9 &0.43 &0.14 &0 &0.501 & {\bf 0.580} &{\bf 0.604} &{\bf 0.625}\\
13   &{\bf 0.32}&{\bf 0.136}&{\bf 0.011}&0.07&{\bf 0.091}&{\bf 0.1197}&{\bf 0.1395} &0.07 &{\bf 0.096}&{\bf 0.125}&{\bf 0.116} &0.9 &0.43 &0.07 &0 &0.501 & {\bf 0.546} &{\bf 0.586} &0.596\\
14   &0.21 &0.134 & 0.000 & 0.07 & 0.081 & {\bf 0.1028} & 0.0375 & 0.07 & 0.075 & 0.100 & 0.055 & 0.9 &{\bf 0.50} &0.14 &0 &0.501 & 0.541 &0.569 &0.584\\
15   &{\bf 0.32}&{\bf 0.136}&{\bf 0.011}&0.07&{\bf 0.091}&{\bf 0.1197}&{\bf 0.1395} &0.07 &{\bf 0.096}&{\bf 0.125}&{\bf 0.116} &0.9 &0.43 &0.07 &0 &0.501 & {\bf 0.546} &{\bf 0.586} &0.596\\
16   &0.00 &0.050 &0.000 &0.00 &0.000 &0.0000 &0.0000 &0.00 &0.000 &0.000 &0.000 &0.0 &0.00 &0.00 &0 &0.500 &0.500 &0.500 &0.500 \\
17   &0.16 &0.133 &0.000 &0.07 &0.074 &0.0538 &0.0205 &0.07 &0.069 &0.050 &0.035 &0.9 &{\bf 0.50} &0.14 &0 &0.501 & 0.536 &0.555 &0.582\\
18   &0.00 &0.050 &0.000 &0.00 &0.000 &0.0000 &0.0000 &0.00 &0.000 &0.000 &0.000 &0.0 &0.00 &0.00 &0 &0.500 &0.500 &0.500 &0.500 \\
19   &0.00 &0.050 &0.000 &0.00 &0.000 &0.0000 &0.0000 &0.00 &0.000 &0.000 &0.000 &0.0 &0.00 &0.00 &0 &0.500 &0.500 &0.500 &0.500 \\
20   &0.00 &0.050 &0.000 &0.00 &0.000 &0.0000 &0.0000 &0.00 &0.000 &0.000 &0.000 &0.0 &0.00 &0.00 &0 &0.500 &0.500 &0.500 &0.500 \\\hline
\end{tabular}
}
\caption{Comparison of centrality measures for the Monkeys' interaction network.
The three more central values of some measures are highlighted in bold. We consider a quota $q=14$.\label{tab:tab1}}
\end{center}
\end{table*}

\begin{flushleft}
\begin{figure}[ht]
\begin{center}
\scalebox{0.7}{
\begin{tikzpicture}
  \begin{axis}[ymin=0, ymax=0.3, xmin=1, xmax=20, height=7cm, width=11cm, grid=major]
    \addplot coordinates {
			(1,0.134)  (2,0.050)  (3,0.143)  (4,0.133)  (5,0.132)  (6,0.050)  (7,0.133)  (8,0.133)  (9,0.131)
			(10,0.133) (11,0.132) (12,0.139) (13,0.136) (14,0.134) (15,0.136) (16,0.050) (17,0.133) (18,0.050)
			(19,0.050) (20,0.050) };
    \addlegendentry{$C_C$}
    \addplot coordinates {
			(1,0.006)  (2,0.000)  (3,0.260)  (4,0.000)  (5,0.000)  (6,0.000)  (7,0.000)  (8,0.003)  (9,0.000)
			(10,0.000) (11,0.000) (12,0.060) (13,0.011) (14,0.000) (15,0.011) (16,0.000) (17,0.000) (18,0.000)
			(19,0.000) (20,0.000) };
    \addlegendentry{$C_B$}
    \addplot[mark=triangle*,color=brown] coordinates {
			(1,0.0885)  (2,0.0000)  (3,0.1730)  (4,0.0343)  (5,0.0438)  (6,0.0000)  (7,0.0460)  (8,0.0863)  (9,0.0003)
			(10,0.0205) (11,0.0035) (12,0.1671) (13,0.1395) (14,0.0375) (15,0.1395) (16,0.0000) (17,0.0205) (18,0.0000)
			(19,0.0000) (20,0.0000) };
    \addlegendentry{{\tt Bz}-C4}
    \addplot coordinates {
			(1,0.075) (2,0.000)   (3,0.192)  (4,0.044)  (5,0.036)  (6,0.000)  (7,0.045)  (8,0.066)  (9,0.005)
			(10,0.035) (11,0.016) (12,0.160) (13,0.116) (14,0.055) (15,0.116) (16,0.000) (17,0.035) (18,0.000)
			(19,0.000) (20,0.000) };
    \addlegendentry{{\tt SS}-C4}
  \end{axis}
	\node[below=0.8cm] at (4.5,.4) {Monkeys};
	\node[rotate=90, above=0.8cm] at (-.7,2.4) {Ranks};
\end{tikzpicture}
}
\caption{Similarities between {\tt Bz}-C4, {\tt SS}-C4, $C_C$ and $C_B$ measures for Monkeys' interaction network.\label{fig:chart_monkeys}}
\end{center}
\end{figure}
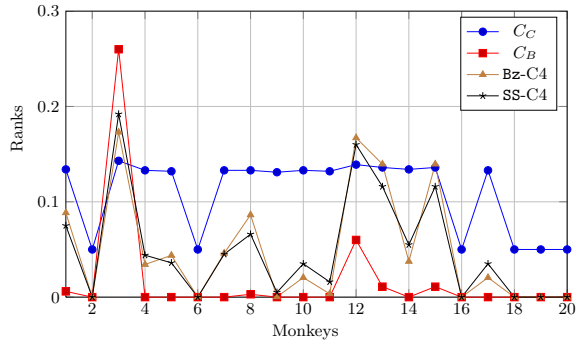
\end{flushleft}

\subsection{Dining-table partners}
\label{sec:Dtp}

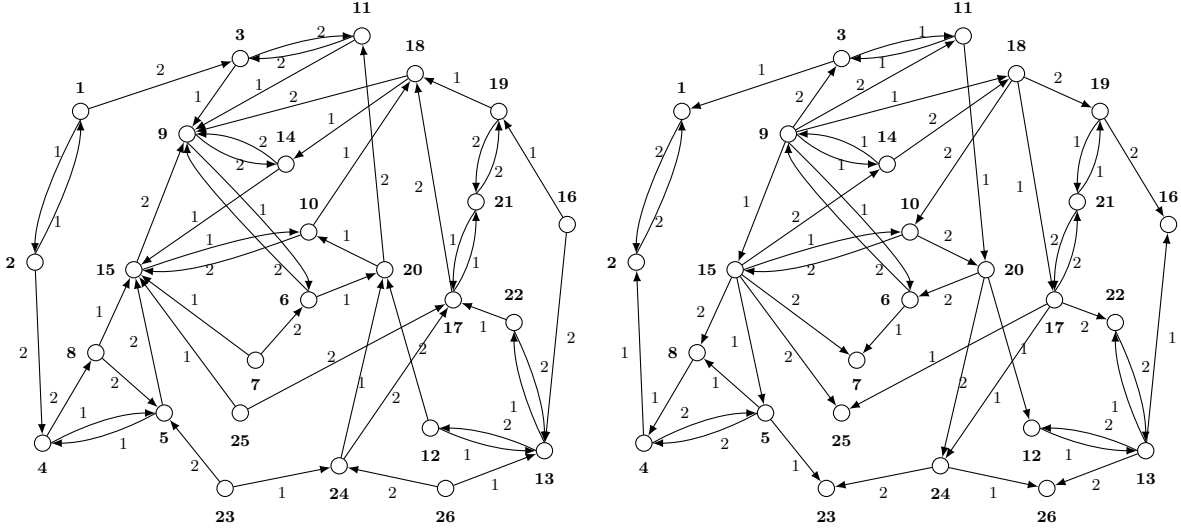
\begin{figure}[ht]
\begin{minipage}[b]{0.49\linewidth}
\begin{center}
\begin{tikzpicture}[every node/.style={circle,scale=0.7}, >=latex]
\node[draw](Ada)     at (0.6,5.0)[label=above:{\bf 1}] {};
\node[draw](Cora)    at (0.0,3.0)[label=left :{\bf 2}] {};
\node[draw](Louise)  at (2.7,5.7)[label=above:{\bf 3}] {};
\node[draw](Jean)    at (0.1,0.6)[label=below:{\bf 4}] {};
\node[draw](Helen)   at (1.7,1.0)[label=below:{\bf 5}] {};
\node[draw](Martha)  at (3.6,2.5)[label=left :{\bf 6}] {};
\node[draw](Alice)   at (2.9,1.7)[label=below:{\bf 7}] {};
\node[draw](Robin)   at (0.8,1.8)[label=left :{\bf 8}] {};
\node[draw](Marion)  at (2.0,4.7)[label=left :{\bf 9}] {};
\node[draw](Maxine)  at (3.6,3.4)[label=above:{\bf 10}]{};
\node[draw](Lena)    at (4.3,6.0)[label=above:{\bf 11}]{};
\node[draw](Hazel)   at (5.2,0.8)[label=below:{\bf 12}]{};
\node[draw](Hilda)   at (6.7,0.5)[label=below:{\bf 13}]{};
\node[draw](Frances) at (3.3,4.3)[label=above:{\bf 14}]{};
\node[draw](Eva)     at (1.3,2.9)[label=left :{\bf 15}]{};
\node[draw](Ruth)    at (7.0,3.5)[label=above:{\bf 16}]{};
\node[draw](Edna)    at (5.5,2.5)[label=below:{\bf 17}]{};
\node[draw](Adele)   at (5.0,5.5)[label=above:{\bf 18}]{};
\node[draw](Jane)    at (6.1,5.0)[label=above:{\bf 19}]{};
\node[draw](Anna)    at (4.6,2.9)[label=right:{\bf 20}]{};
\node[draw](Mary)    at (5.8,3.8)[label=right:{\bf 21}]{};
\node[draw](Betty)   at (6.3,2.2)[label=above:{\bf 22}]{};
\node[draw](Ella)    at (2.5,0.0)[label=below:{\bf 23}]{};
\node[draw](Ellen)   at (4.0,0.3)[label=below:{\bf 24}]{};
\node[draw](Laura)   at (2.7,1.0)[label=below:{\bf 25}]{};
\node[draw](Irene)   at (5.4,0.0)[label=below:{\bf 26}]{};
\draw[->] (Ada)     .. controls +(down:.1cm) and +(up:.8cm)    .. node[above]{1}(Cora);
\draw[->] (Ada)                                                to node[above]{2}(Louise);
\draw[->] (Cora)    .. controls +(up:.1cm) and +(down:.8cm)    .. node[below]{1}(Ada);
\draw[->] (Cora)                                               to node[left] {2}(Jean);
\draw[->] (Louise)                                             to node[left] {1}(Marion);
\draw[->] (Louise)  .. controls +(right:.1cm) and +(left:.8cm) .. node[below]{2}(Lena);
\draw[->] (Jean)    .. controls +(right:.1cm) and +(left:.8cm) .. node[above]{1}(Helen);
\draw[->] (Jean)                                               to node[left] {2}(Robin);
\draw[->] (Helen)   .. controls +(left:.1cm) and +(right:.8cm) .. node[below]{1}(Jean);
\draw[->] (Helen)                                              to node[left] {2}(Eva);
\draw[->] (Martha)  					       to node[below]{1}(Anna);
\draw[->] (Martha)  .. controls +(up:.1cm) and +(down:.5cm)    .. node[below]{ }(Marion);
\node[scale=1]      at (3.2,2.9)					     {2};
\draw[->] (Alice)   					       to node[above]{1}(Eva);
\draw[->] (Alice)   					       to node[right]{2}(Martha);
\draw[->] (Robin)                                              to node[left] {1}(Eva);
\draw[->] (Robin)                                              to node[left] {2}(Helen);
\draw[->] (Marion)  .. controls +(down:.1cm) and +(up:.5cm)    .. node[right]{ }(Martha);
\node[scale=1]      at (3.0,3.7)					     {1};
\draw[->] (Marion)  .. controls +(right:.1cm) and +(left:.8cm) .. node[right]{ }(Frances);
\node[scale=1]      at (2.7,4.3)					     {2};
\draw[->] (Maxine)  					       to node[left] {1}(Adele);
\draw     (Maxine)  edge[out=-160, in=-5, looseness=0.8, ->]      node[below]{ }(Eva);
\node[scale=1]      at (2.3,2.9)					     {2};
\draw[->] (Lena)    					       to node[left] {1}(Marion);
\draw[->] (Lena)    .. controls +(left:.1cm) and +(right:.8cm) .. node[above]{2}(Louise);
\draw[->] (Hazel)   .. controls +(right:.1cm) and +(left:.8cm) .. node[below]{1}(Hilda);
\draw[->] (Hazel)   					       to node[right]{2}(Anna);
\draw[->] (Hilda)   .. controls +(up:.1cm) and +(down:.8cm)    .. node[left] {1}(Betty);
\draw[->] (Hilda)   .. controls +(left:.1cm) and +(right:.8cm) .. node[above]{2}(Hazel);
\draw[->] (Frances) 					       to node[above]{ }(Eva);
\node[scale=1]      at (2.1,3.6)					     {1};
\draw[->] (Frances) .. controls +(left:.1cm) and +(right:.8cm) .. node[above]{ }(Marion);
\node[scale=1]      at (3.0,4.6)					     {2};
\draw[->] (Eva)     .. controls +(right:.1cm) and +(left:.8cm) .. node[above]{ }(Maxine);
\node[scale=1]      at (2.3,3.3)					     {1};
\draw[->] (Eva)     					       to node[left] {2}(Marion);
\draw[->] (Ruth)    					       to node[above]{1}(Jane);
\draw[->] (Ruth)    					       to node[right]{2}(Hilda);
\draw[->] (Edna)    .. controls +(up:.1cm) and +(down:.8cm)    .. node[right]{ }(Mary);
\node[scale=1]      at (5.8,3.0)					     {1};
\draw[->] (Edna)    					       to node[left] {2}(Adele);
\draw[->] (Adele)   					       to node[above]{ }(Frances);
\node[scale=1]      at (3.9,4.9)					     {1};
\draw[->] (Adele)   					       to node[left] { }(Marion);
\node[scale=1]      at (3.4,5.2)					     {2};
\draw[->] (Jane)    					       to node[above]{1}(Adele);
\draw[->] (Jane)    .. controls +(down:.1cm) and +(up:.8cm)    .. node[left] { }(Mary);
\node[scale=1]      at (5.8,4.6)					     {2};
\draw[->] (Anna)    					       to node[above]{1}(Maxine);
\draw[->] (Anna)    					       to node[right]{ }(Lena);
\node[scale=1]      at (4.6,4.1)					     {2};
\draw[->] (Mary)    .. controls +(down:.1cm) and +(up:.8cm)    .. node[left] { }(Edna);
\node[scale=1]      at (5.5,3.3)					     {1};
\draw[->] (Mary)    .. controls +(up:.1cm) and +(down:.8cm)    .. node[right]{ }(Jane);
\node[scale=1]      at (6.1,4.2)					     {2};
\draw[->] (Betty)   					       to node[below]{1}(Edna);
\draw[->] (Betty)   .. controls +(down:.1cm) and +(up:.8cm)    .. node[right]{2}(Hilda);
\draw[->] (Ella)    					       to node[below]{1}(Ellen);
\draw[->] (Ella)    					       to node[below]{2}(Helen);
\draw[->] (Ellen)   					       to node[below]{1}(Anna);
\draw[->] (Ellen)   					       to node[below]{2}(Edna);
\draw[->] (Laura)   					       to node[below]{1}(Eva);
\draw[->] (Laura)   					       to node[left] {2}(Edna);
\draw[->] (Irene)   					       to node[below]{1}(Hilda);
\draw[->] (Irene)   					       to node[below]{2}(Ellen);
\end{tikzpicture}
\end{center}
\end{minipage}
\begin{minipage}[b]{0.49\linewidth}
\begin{center}
\begin{tikzpicture}[every node/.style={circle,scale=0.7}, >=latex]
\node[draw](Ada)     at (0.6,5.0)[label=above:{\bf 1}] {};
\node[draw](Cora)    at (0.0,3.0)[label=left :{\bf 2}] {};
\node[draw](Louise)  at (2.7,5.7)[label=above:{\bf 3}] {};
\node[draw](Jean)    at (0.1,0.6)[label=below:{\bf 4}] {};
\node[draw](Helen)   at (1.7,1.0)[label=below:{\bf 5}] {};
\node[draw](Martha)  at (3.6,2.5)[label=left :{\bf 6}] {};
\node[draw](Alice)   at (2.9,1.7)[label=below:{\bf 7}] {};
\node[draw](Robin)   at (0.8,1.8)[label=left :{\bf 8}] {};
\node[draw](Marion)  at (2.0,4.7)[label=left :{\bf 9}] {};
\node[draw](Maxine)  at (3.6,3.4)[label=above:{\bf 10}]{};
\node[draw](Lena)    at (4.3,6.0)[label=above:{\bf 11}]{};
\node[draw](Hazel)   at (5.2,0.8)[label=below:{\bf 12}]{};
\node[draw](Hilda)   at (6.7,0.5)[label=below:{\bf 13}]{};
\node[draw](Frances) at (3.3,4.3)[label=above:{\bf 14}]{};
\node[draw](Eva)     at (1.3,2.9)[label=left :{\bf 15}]{};
\node[draw](Ruth)    at (7.0,3.5)[label=above:{\bf 16}]{};
\node[draw](Edna)    at (5.5,2.5)[label=below:{\bf 17}]{};
\node[draw](Adele)   at (5.0,5.5)[label=above:{\bf 18}]{};
\node[draw](Jane)    at (6.1,5.0)[label=above:{\bf 19}]{};
\node[draw](Anna)    at (4.6,2.9)[label=right:{\bf 20}]{};
\node[draw](Mary)    at (5.8,3.8)[label=right:{\bf 21}]{};
\node[draw](Betty)   at (6.3,2.2)[label=above:{\bf 22}]{};
\node[draw](Ella)    at (2.5,0.0)[label=below:{\bf 23}]{};
\node[draw](Ellen)   at (4.0,0.3)[label=below:{\bf 24}]{};
\node[draw](Laura)   at (2.7,1.0)[label=below:{\bf 25}]{};
\node[draw](Irene)   at (5.4,0.0)[label=below:{\bf 26}]{};
\draw[->] (Ada)     .. controls +(down:.1cm) and +(up:.8cm)    .. node[above]{2}(Cora);
\draw[->] (Louise)                                             to node[above]{1}(Ada);
\draw[->] (Cora)    .. controls +(up:.1cm) and +(down:.8cm)    .. node[below]{2}(Ada);
\draw[->] (Jean)                                               to node[left] {1}(Cora);
\draw[->] (Marion)                                             to node[left] {2}(Louise);
\draw[->] (Louise)  .. controls +(right:.1cm) and +(left:.8cm) .. node[below]{1}(Lena);
\draw[->] (Jean)    .. controls +(right:.1cm) and +(left:.8cm) .. node[above]{2}(Helen);
\draw[->] (Robin)                                              to node[left] {1}(Jean);
\draw[->] (Helen)   .. controls +(left:.1cm) and +(right:.8cm) .. node[below]{2}(Jean);
\draw[->] (Eva)                                                to node[left] {1}(Helen);
\draw[->] (Anna)  					       to node[below]{2}(Martha);
\draw[->] (Martha)  .. controls +(up:.1cm) and +(down:.5cm)    .. node[below]{ }(Marion);
\node[scale=1]      at (3.2,2.9)					     {2};
\draw[->] (Eva)   					       to node[above]{2}(Alice);
\draw[->] (Martha)   					       to node[right]{1}(Alice);
\draw[->] (Eva)                                                to node[left] {2}(Robin);
\draw[->] (Helen)                                              to node[left] {1}(Robin);
\draw[->] (Marion)  .. controls +(down:.1cm) and +(up:.5cm)    .. node[right]{ }(Martha);
\node[scale=1]      at (3.0,3.7)					     {1};
\draw[->] (Marion)  .. controls +(right:.1cm) and +(left:.8cm) .. node[right]{ }(Frances);
\node[scale=1]      at (2.7,4.3)					     {1};
\draw[->] (Adele)  					       to node[left] {2}(Maxine);
\draw     (Maxine)  edge[out=-160, in=-5, looseness=0.8, ->]      node[below]{ }(Eva);
\node[scale=1]      at (2.3,2.9)					     {2};
\draw[->] (Marion)    					       to node[left] {2}(Lena);
\draw[->] (Lena)    .. controls +(left:.1cm) and +(right:.8cm) .. node[above]{1}(Louise);
\draw[->] (Hazel)   .. controls +(right:.1cm) and +(left:.8cm) .. node[below]{1}(Hilda);
\draw[->] (Anna)   					       to node[right]{1}(Hazel);
\draw[->] (Hilda)   .. controls +(up:.1cm) and +(down:.8cm)    .. node[left] {1}(Betty);
\draw[->] (Hilda)   .. controls +(left:.1cm) and +(right:.8cm) .. node[above]{2}(Hazel);
\draw[->] (Eva) 					       to node[above]{ }(Frances);
\node[scale=1]      at (2.1,3.6)					     {2};
\draw[->] (Frances) .. controls +(left:.1cm) and +(right:.8cm) .. node[above]{ }(Marion);
\node[scale=1]      at (3.0,4.6)					     {1};
\draw[->] (Eva)     .. controls +(right:.1cm) and +(left:.8cm) .. node[above]{ }(Maxine);
\node[scale=1]      at (2.3,3.3)					     {1};
\draw[->] (Marion)     					       to node[left] {1}(Eva);
\draw[->] (Jane)    					       to node[above]{2}(Ruth);
\draw[->] (Hilda)    					       to node[right]{1}(Ruth);
\draw[->] (Edna)    .. controls +(up:.1cm) and +(down:.8cm)    .. node[right]{ }(Mary);
\node[scale=1]      at (5.8,3.0)					     {2};
\draw[->] (Adele)    					       to node[left] {1}(Edna);
\draw[->] (Frances)   					       to node[above]{ }(Adele);
\node[scale=1]      at (3.9,4.9)					     {2};
\draw[->] (Marion)   					       to node[left] { }(Adele);
\node[scale=1]      at (3.4,5.2)					     {1};
\draw[->] (Adele)    					       to node[above]{2}(Jane);
\draw[->] (Jane)    .. controls +(down:.1cm) and +(up:.8cm)    .. node[left] { }(Mary);
\node[scale=1]      at (5.8,4.6)					     {1};
\draw[->] (Maxine)    					       to node[above]{2}(Anna);
\draw[->] (Lena)    					       to node[right]{ }(Anna);
\node[scale=1]      at (4.6,4.1)					     {1};
\draw[->] (Mary)    .. controls +(down:.1cm) and +(up:.8cm)    .. node[left] { }(Edna);
\node[scale=1]      at (5.5,3.3)					     {2};
\draw[->] (Mary)    .. controls +(up:.1cm) and +(down:.8cm)    .. node[right]{ }(Jane);
\node[scale=1]      at (6.1,4.2)					     {1};
\draw[->] (Edna)   					       to node[below]{2}(Betty);
\draw[->] (Betty)   .. controls +(down:.1cm) and +(up:.8cm)    .. node[right]{2}(Hilda);
\draw[->] (Ellen)    					       to node[below]{2}(Ella);
\draw[->] (Helen)    					       to node[below]{1}(Ella);
\draw[->] (Anna)   					       to node[below]{2}(Ellen);
\draw[->] (Edna)   					       to node[below]{1}(Ellen);
\draw[->] (Eva)   					       to node[below]{2}(Laura);
\draw[->] (Edna)   					       to node[left] {1}(Laura);
\draw[->] (Hilda)   					       to node[below]{2}(Irene);
\draw[->] (Ellen)   					       to node[below]{1}(Irene);
\end{tikzpicture}
\end{center}
\end{minipage}
\caption{Original network for dining-table partners \cite{Mor60,dNMB04}, and the adaptation to influence graph.\label{fig3}}
\end{figure}

A second real network is illustrated in Figure~\ref{fig3}. It was firstly provided by a sociometric research of \cite{Mor60} and,
years later, it was also used by \cite{dNMB04} to be handled and displayed by a computational application.

It represents the companion preferences of 26 girls living in one cottage at a New York state training school.
Each girl was asked about who prefers as dining-table partner in first and second place.
Thus, each girl is represented by a node, and there is a directed edge $(i,j)$ per each girl $i$ prefering girl $j$ as dining-table partner.
Every node has an outdegree equal to 2: edges with weight 1 denote the first option of the girl, and edges with weight 2 denote her second option.

We could assume that a girl has some ability to influence over another one which has chosen her as a partner.
Figure~\ref{fig3} also shows the corresponding network of this influence game,
reversing each arc $(i,j)$ by $(j,i)$, so that a node points to another when the first one has some influence over the second one. Further, the weights of the edges must be exchanged, so that an original edge $(i,j)$ with weight 1 now becomes in an edge $(j,i)$ with weight 2, and viceversa.
This is due to a girl has more influence over another one if that other has chosen her in first place rather than in second place.
Of course, now every node has an indegree equal to 2: one edge with weight 1 and the other with weight~2.

Instead of the Monkeys' interaction network, here there is no isolated nodes, but we can still obtaining scores for {\tt Bz} and {\tt SS} measures equal to zero. For instance, see the columns of {\tt Bz}-C1 and {\tt SS}-C1 on Table~\ref{tab:tab2}.

A common voting system is the one of {\em absolute majority}, in which an option wins whether it has more than the half of the votes.
According to this idea, we consider for our experiments a quota $q=14$, so that a coalition is considered successful or winning if and only if through its spread of influence, this coalition achieves to convince most of the girls.

For every node $i\in V$, we consider the following reasonable labeling functions:
\begin{itemize}
 \item Case 1 (C1): Minimum influence required to convincement, $f(i)=1$.
 \item Case 2 (C2): Average influence required to convincement, $f(i)=2$.
 \item Case 3 (C3): Maximum influence required to convincement, $f(i)=3$.
\end{itemize}
The comparison between traditional measures and the {\tt Bz}, {\tt SS}, $C_E$ and $C_S$ measures are shown on Table~\ref{tab:tab2}.

\begin{table*}[ht]
\begin{center}
\scalebox{0.65}{
\begin{tabular}{|c|c|c|c|c|c|c|c|c|c|c|c|c|c|c|c|c|}\cline{6-17}
\multicolumn{5}{c|}{\,} & \multicolumn{3}{c|}{\tt Bz} & \multicolumn{3}{c|}{\tt SS} & \multicolumn{3}{c|}{$C_E$} & \multicolumn{3}{c|}{$C_S$}\\\hline
Node & $C_D^-$ & $C_D^+$ & $C_C$ & $C_B$ & C1 & C2 & C3 & C1 & C2 & C3 & C1 & C2 & C3 & C1 & C2 & C3 \\\hline
1  & 0.08 & 0.04 & 0.0400 & 0.035 & 0.00 & 0.028 & 0.0274 & 0.0000 & 0.0103 & 0.0259 & 0.92 & 0.85 & 0.42 & 0.500000 & 0.5024 & 0.5300\\
2  & 0.08 & 0.04 & 0.0400 & 0.033 & 0.00 & 0.028 & 0.0274 & 0.0000 & 0.0103 & 0.0259 & 0.92 & 0.85 & 0.42 & 0.500000 & 0.5024 & 0.5300\\
3  & 0.08 & 0.08 & 0.2273 & 0.072 & {\bf 0.08} & 0.008 & 0.0302 & {\bf 0.0832} & 0.0014 & 0.0331 & {\bf 0.96} & 0.85 & {\bf 0.54} & {\bf 0.500217} & 0.5006 & 0.5329\\
4  & 0.08 & 0.08 & 0.0473 & 0.039 & 0.00 & 0.028 & 0.0413 & 0.0000 & 0.0103 & 0.0383 & 0.92 & 0.85 & 0.42 & 0.500000 & 0.5024 & 0.5451\\
5  & 0.08 & 0.12 & 0.0473 & 0.049 & 0.00 & 0.028 & 0.0452 & 0.0000 & 0.0103 & 0.0463 & 0.92 & 0.85 & 0.42 & 0.500000 & 0.5024 & 0.5494\\
6  & 0.08 & 0.08 & 0.3165 & 0.102 & {\bf 0.08} & 0.043 & 0.0481 & {\bf 0.0832} & 0.0142 & 0.0473 & {\bf 0.96} & 0.85 & 0.42 & {\bf 0.500217} & 0.5036 & 0.5526\\
7  & 0.08 & 0.00 & 0.0385 & 0.015 & 0.01 & 0.024 & 0.0216 & 0.0003 & 0.0075 & 0.0176 & 0.92 & 0.85 & 0.42 & 0.500027 & 0.5020 & 0.5236\\
8  & 0.08 & 0.04 & 0.0471 & 0.036 & 0.00 & 0.024 & 0.0278 & 0.0000 & 0.0075 & 0.0239 & 0.92 & 0.85 & 0.42 & 0.500000 & 0.5020 & 0.5303\\
9  & 0.08 & {\bf 0.24} & {\bf 0.4902} & {\bf 0.232} & {\bf 0.08} & 0.027 & {\bf 0.0820} & {\bf 0.0832} & 0.0072 & {\bf 0.0965} & {\bf 0.96} & 0.85 & {\bf 0.54} & {\bf 0.500217} & 0.5023 & {\bf 0.5896}\\
10 & 0.08 & 0.08 & 0.3378 & 0.089 & {\bf 0.08} & {\bf 0.104} & {\bf 0.0506} & {\bf 0.0832} & {\bf 0.1953} & {\bf 0.0578} & {\bf 0.96} & {\bf 0.92} & {\bf 0.54} & {\bf 0.500217} & {\bf 0.5089} & {\bf 0.5552}\\
11 & 0.08 & 0.08 & 0.2778 & 0.107 & {\bf 0.08} & 0.008 & 0.0383 & {\bf 0.0832} & 0.0014 & 0.0410 & {\bf 0.96} & 0.85 & {\bf 0.54} & {\bf 0.500217} & 0.5006 & 0.5418\\
12 & 0.08 & 0.04 & 0.0452 & 0.052 & 0.00 & 0.004 & 0.0321 & 0.0001 & 0.0007 & 0.0292 & 0.92 & 0.85 & 0.42 & 0.500004 & 0.5003 & 0.5350\\
13 & 0.08 & {\bf 0.16} & 0.0454 & 0.061 & 0.00 & 0.014 & 0.0500	& 0.0001 & 0.0041 & 0.0511 & 0.92 & 0.85 & 0.42 & 0.500004 & 0.5012 & 0.5546\\
14 & 0.08 & 0.08 & {\bf 0.3571} & 0.083 & {\bf 0.08} & {\bf 0.104} & 0.0486 & {\bf 0.0832} & {\bf 0.1953} & 0.0465 & {\bf 0.96} & {\bf 0.92} & 0.42 & {\bf 0.500217} & {\bf 0.5089} & 0.5531\\
15 & 0.08 & {\bf 0.24} & {\bf 0.3906} & {\bf 0.145} & {\bf 0.08} & {\bf 0.104} & {\bf 0.0683} & {\bf 0.0832} &{\bf 0.1953}&{\bf 0.0755}&{\bf 0.96}&{\bf 0.92}&{\bf 0.54} & {\bf 0.500217} & {\bf 0.5089} & {\bf 0.5746}\\
16 & 0.08 & 0.00 & 0.0385 & 0.025 & 0.00 & 0.015 & 0.0279 & 0.0000 & 0.0055 & 0.0256 & 0.92 & 0.85 & 0.42 & 0.500000 & 0.5013 & 0.5305\\
17 & 0.08 & {\bf 0.33} & 0.0614 & 0.091 & {\bf 0.08} & {\bf 0.051} & 0.0469 & {\bf 0.0832} & {\bf 0.0232} & 0.0459 & {\bf 0.96} & 0.85 & 0.42 & {\bf 0.500217} & {\bf 0.5043} & 0.5512\\
18 & 0.08 & 0.12 & 0.3425 & 0.123 & {\bf 0.08} & {\bf 0.104} & 0.0404 & {\bf 0.0832} & {\bf 0.1953} & 0.0361 & {\bf 0.96} & {\bf 0.92} & 0.42 & {\bf 0.500217} & {\bf 0.5089} & 0.5442\\
19 & 0.08 & 0.08 & 0.0595 & 0.053 & {\bf 0.08} & 0.036 & 0.0356 & {\bf 0.0832} & 0.0126 & 0.0327 & {\bf 0.96} & 0.85 & 0.42 & {\bf 0.500217} & 0.5031 & 0.5389\\
20 & 0.08 & 0.12 & 0.3247 & {\bf 0.164} & {\bf 0.08} & {\bf 0.075} & 0.0394 & {\bf 0.0832} & {\bf 0.0413} & 0.0395 & {\bf 0.96} & 0.85 & 0.42 & {\bf 0.500217} & {\bf 0.5064} & 0.5430\\
21 & 0.08 & 0.08 & 0.0605 & 0.038 & {\bf 0.08} & {\bf 0.051} & 0.0457 & {\bf 0.0832} & {\bf 0.0232} & 0.0512 & {\bf 0.96} & 0.85 & {\bf 0.54} & {\bf 0.500217} & {\bf 0.5043} & 0.5499\\
22 & 0.08 & 0.04 & 0.0452 & 0.046 & 0.00 & 0.025 & 0.0325 & 0.0001 & 0.0082 & 0.0293 & 0.92 & 0.85 & 0.42 & 0.500004 & 0.5021 & 0.5355\\
23 & 0.08 & 0.00 & 0.0385 & 0.027 & 0.00 & 0.011 & 0.0191 & 0.0000 & 0.0029 & 0.0173 & 0.92 & 0.85 & 0.42 & 0.500000 & 0.5010 & 0.5208\\
24 & 0.08 & 0.08 & 0.0417 & 0.057 & 0.01 & 0.029 & 0.0324 & 0.0003 & 0.0083 & 0.0301 & 0.92 & 0.85 & 0.42 & 0.500027 & 0.5025 & 0.5354\\
25 & 0.08 & 0.00 & 0.0385 & 0.020 & 0.01 & 0.024 & 0.0218 & 0.0003 & 0.0075 & 0.0187 & 0.92 & 0.85 & 0.42 & 0.500027 & 0.5020 & 0.5239\\
26 & 0.08 & 0.00 & 0.0385 & 0.027 & 0.00 & 0.004 & 0.0197 & 0.0000 & 0.0007 & 0.0177 & 0.92 & 0.85 & 0.42 & 0.500000 & 0.5003 & 0.5215\\\hline
\end{tabular}
}
\caption{Comparison of centrality measures for the influence game version of the Dining-table partners network.
The three more central values of some measures are highlighted in bold. We consider a quota $q=14$.\label{tab:tab2}}
\end{center}
\end{table*}

Note that in this case, indegree centrality $C_D^-$ does not provide any relevant information, because the indegree for each node is always~2.

Similarly as it succeded in previous section, {\tt Bz}-C1, {\tt SS}-C1, $C_E$-C1 and $C_S$-C1 have a lot of nodes with the same rank,
but while increases the required influence to convincement, the values of the measures are more diverse for the power indices and satisfaction centrality.
Note that measures {\tt Bz}-C2, {\tt SS}-C2 and $C_S$-C2, as well as $C_D^+$ and $C_C$, have only some values ​​that are repeated,
but measures {\tt Bz}-C3, {\tt SS}-C3 and $C_S$-C3 have the same values only for girls 1 and 2. These girls are equivalent in this sense for all the other measures except by $C_B$, in which, however, together with $C_E$, girls 23 and 26 have the same centrality.

The most central girls are highlighted in Table \ref{tab:tab2}. Girl 15 has a high centrality in all measures, as well as girl 9, except in $C_E$-C2, as well as in {\tt Bz}-C2 and {\tt SS}-C2, where is far less central. Note that girl 13 is fairly central exclusively in $C_D^+$, because despite of its high outdegree, only exist paths from this node to another four, which is a severe restriction for all other measures considered.

Additionally, unlike traditional measures, girl 10 play an important role in our new measures because,
in spite of neither having a high outdegree nor having too short paths to more distant nodes,
she plays an essential role in the spread of influence to convince distant sets of girls, which in turn have no convincing power over her.

Finally, note that as in the previous network ---see Figure~\ref{fig:chart}--- measures {\tt Bz} and {\tt SS} produce similar rankings.
Moreover, it will be also fulfilled on the network of the next section.
Otherwise, on the next network the measure $C_E$ will produce more varied values than in this network and the previous one.

\begin{flushleft}
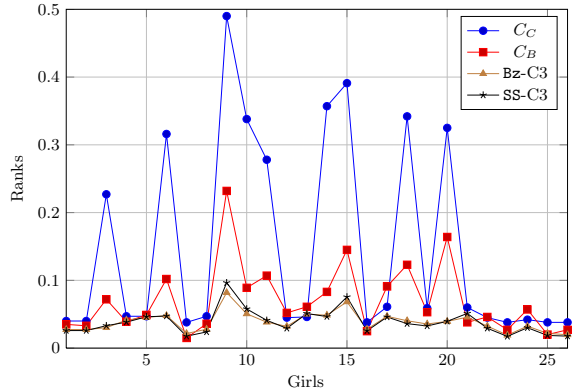
\begin{figure}[ht]
\begin{center}
\scalebox{0.7}{
\begin{tikzpicture}
  \begin{axis}[ymin=0, ymax=0.5, xmin=1, xmax=26, height=8cm, width=11cm, grid=major]
    \addplot coordinates {
			(1,0.040)  (2,0.040)  (3,0.227)  (4,0.047)  (5,0.047)  (6,0.316)  (7,0.038)  (8,0.047)  (9,0.490)
			(10,0.338) (11,0.278) (12,0.045) (13,0.046) (14,0.357) (15,0.391) (16,0.038) (17,0.061) (18,0.342)
			(19,0.059) (20,0.325) (21,0.060) (22,0.045) (23,0.038) (24,0.042) (25,0.038) (26,0.038) };
    \addlegendentry{$C_C$}
    \addplot coordinates {
			(1,0.035)  (2,0.033)  (3,0.072)  (4,0.039)  (5,0.049)  (6,0.102)  (7,0.015)  (8,0.036)  (9,0.232)
			(10,0.089) (11,0.107) (12,0.052) (13,0.061) (14,0.083) (15,0.145) (16,0.025) (17,0.091) (18,0.123)
			(19,0.053) (20,0.164) (21,0.038) (22,0.046) (23,0.027) (24,0.057) (25,0.020) (26,0.027) };
    \addlegendentry{$C_B$}
    \addplot[mark=triangle*,color=brown] coordinates {
			(1,0.0274)  (2,0.0274)  (3,0.0302)  (4,0.0413)  (5,0.0452)  (6,0.0481)  (7,0.0216)  (8,0.0278)  (9,0.0820)
			(10,0.0506) (11,0.0383) (12,0.0321) (13,0.0500) (14,0.0486) (15,0.0683) (16,0.0279) (17,0.0469) (18,0.0404)
			(19,0.0356) (20,0.0394) (21,0.0457) (22,0.0325) (23,0.0191) (24,0.0324) (25,0.0218) (26,0.0197) };
    \addlegendentry{{\tt Bz}-C3}
    \addplot coordinates {
			(1,0.0259) 	(2,0.0259) 	(3,0.0331) 	(4,0.0383)  (5,0.0463)  (6,0.0473)  (7,0.0176)  (8,0.0239)  (9,0.0965)
			(10,0.0578) (11,0.0410) (12,0.0292) (13,0.0511) (14,0.0465) (15,0.0755) (16,0.0256) (17,0.0459) (18,0.0361)
			(19,0.0327) (20,0.0395) (21,0.0512) (22,0.0293) (23,0.0173) (24,0.0301) (25,0.0187) (26,0.0177) };
    \addlegendentry{{\tt SS}-C3}
  \end{axis}
	\node[below=0.8cm] at (4.5,.4) {Girls};
	\node[rotate=90, above=0.8cm] at (-.7,2.4) {Ranks};
\end{tikzpicture}
}
\caption{Similarities between {\tt Bz}-C3, {\tt SS}-C3, $C_C$ and $C_B$ measures for Dining-table partners network.\label{fig:chart}}
\end{center}
\end{figure}
\end{flushleft}

\subsection{Student Government discussion}
\label{sec:SGd}

\begin{figure}[ht]
\begin{minipage}[b]{0.49\linewidth}
\begin{center}
\begin{tikzpicture}[every node/.style={circle,scale=0.7}, >=latex]
\node[draw](1) at (0.0,2.8)[label=left :{\bf 1 }]{2};
\node[draw](2) at (2.3,2.5)[label=above:{\bf 2 }]{3};
\node[draw](3) at (1.0,0.7)[label=left :{\bf 3 }]{2};
\node[draw](4) at (1.6,4.2)[label=above:{\bf 4 }]{2};
\node[draw](5) at (3.2,5.2)[label=above:{\bf 5 }]{2};
\node[draw](6) at (5.8,4.2)[label=above:{\bf 6 }]{2};
\node[draw](7) at (4.5,2.6)[label=above:{\bf 7 }]{2};
\node[draw](8) at (4.4,0.9)[label=below:{\bf 8 }]{2};
\node[draw](9) at (2.2,0.0)[label=left :{\bf 9 }]{1};
\node[draw](10)at (0.4,4.6)[label=left :{\bf 10}]{1};
\node[draw](11)at (5.7,0.1)[label=right:{\bf 11}]{1};
\draw[->] (1) to node {}(2);
\path (1) edge [->,bend left=5] node {}(3);
\draw[->] (1) to node {}(6);
\path (2) edge [->,bend right=5] node {}(8);
\path (3) edge [->,bend left=5] node {}(1);
\draw[->] (3) to node {}(2);
\draw[->] (3) to node {}(4);
\draw[->] (3) to node {}(6);
\draw[->] (3) to node {}(7);
\draw[->] (3) to node {}(8);
\path (4) edge [->,bend left=5] node {}(7);
\path (4) edge [->,bend left=5] node {}(8);
\draw[->] (5) to node {}(2);
\draw[->] (5) to node {}(4);
\path (5) edge [->,bend left=5] node {}(6);
\draw[->] (5) to node {}(7);
\draw[->] (5) to node {}(8);
\draw[->] (6) to node {}(2);
\draw[->] (6) to node {}(4);
\path (6) edge [->,bend left=5] node {}(5);
\draw[->] (6) to node {}(7);
\draw[->] (6) to node {}(8);
\path (7) edge [->,bend left=5] node {}(4);
\path (7) edge [->,bend left=5] node {}(8);
\path (7) edge [->,bend left=5] node {}(9);
\draw[->] (7) to node {}(11);
\path (8) edge [->,bend right=5] node {}(2);
\path (8) edge [->,bend left=5] node {}(4);
\path (8) edge [->,bend left=5] node {}(7);
\path (8) edge [->,bend left=5] node {}(11);
\draw[->] (9) to node {}(4);
\path (9) edge [->,bend left=5] node {}(7);
\draw[->] (9) to node {}(8);
\path (9) edge [->,bend left=5] node {}(11);
\draw[->] (10)to node {}(1);
\draw[->] (10)to node {}(3);
\draw[->] (10)to node {}(4);
\draw[->] (10)to node {}(5);
\draw[->] (11)to node {}(6);
\path (11) edge [->,bend left=5] node {}(8);
\path (11) edge [->,bend left=5] node {}(9);
\end{tikzpicture}
\end{center}
\end{minipage}
\begin{minipage}[b]{0.49\linewidth}
\begin{center}
\begin{tikzpicture}[every node/.style={circle,scale=0.7}, >=latex]
\node[draw](1) at (0.0,2.8)[label=left :{\bf 1 }]{1};
\node[draw](2) at (2.3,2.5)[label=above:{\bf 2 }]{5};
\node[draw](3) at (1.0,0.7)[label=left :{\bf 3 }]{1};
\node[draw](4) at (1.6,4.2)[label=above:{\bf 4 }]{4};
\node[draw](5) at (3.2,5.2)[label=above:{\bf 5 }]{1};
\node[draw](6) at (5.8,4.2)[label=above:{\bf 6 }]{2};
\node[draw](7) at (4.5,2.6)[label=above:{\bf 7 }]{3};
\node[draw](8) at (4.4,0.9)[label=below:{\bf 8 }]{4};
\node[draw](9) at (2.2,0.0)[label=left :{\bf 9 }]{1};
\node[draw](10)at (0.4,4.6)[label=left :{\bf 10}]{1};
\node[draw](11)at (5.7,0.1)[label=right:{\bf 11}]{1};
\draw[->] (1) to node {}(2);
\path (1) edge [->,bend left=5] node {}(3);
\draw[->] (1) to node {}(6);
\path (2) edge [->,bend right=5] node {}(8);
\path (3) edge [->,bend left=5] node {}(1);
\draw[->] (3) to node {}(2);
\draw[->] (3) to node {}(4);
\draw[->] (3) to node {}(6);
\draw[->] (3) to node {}(7);
\draw[->] (3) to node {}(8);
\path (4) edge [->,bend left=5] node {}(7);
\path (4) edge [->,bend left=5] node {}(8);
\draw[->] (5) to node {}(2);
\draw[->] (5) to node {}(4);
\path (5) edge [->,bend left=5] node {}(6);
\draw[->] (5) to node {}(7);
\draw[->] (5) to node {}(8);
\draw[->] (6) to node {}(2);
\draw[->] (6) to node {}(4);
\path (6) edge [->,bend left=5] node {}(5);
\draw[->] (6) to node {}(7);
\draw[->] (6) to node {}(8);
\path (7) edge [->,bend left=5] node {}(4);
\path (7) edge [->,bend left=5] node {}(8);
\path (7) edge [->,bend left=5] node {}(9);
\draw[->] (7) to node {}(11);
\path (8) edge [->,bend right=5] node {}(2);
\path (8) edge [->,bend left=5] node {}(4);
\path (8) edge [->,bend left=5] node {}(7);
\path (8) edge [->,bend left=5] node {}(11);
\draw[->] (9) to node {}(4);
\path (9) edge [->,bend left=5] node {}(7);
\draw[->] (9) to node {}(8);
\path (9) edge [->,bend left=5] node {}(11);
\draw[->] (10)to node {}(1);
\draw[->] (10)to node {}(3);
\draw[->] (10)to node {}(4);
\draw[->] (10)to node {}(5);
\draw[->] (11)to node {}(6);
\path (11) edge [->,bend left=5] node {}(8);
\path (11) edge [->,bend left=5] node {}(9);
\node[scale=1] at (1.6,2.7) {2}; 
\node[scale=1] at (0.4,1.5) {2}; 
\node[scale=1] at (1.1,3.2) {2}; 
\node[scale=1] at (3.2,1.6) {3}; 
\node[scale=1] at (0.6,2.0) {2}; 
\node[scale=1] at (1.6,1.7) {2}; 
\node[scale=1] at (1.2,2.3) {2}; 
\node[scale=1] at (2.4,1.9) {2}; 
\node[scale=1] at (2.7,1.5) {2}; 
\node[scale=1] at (2.5,0.9) {2}; 
\node[scale=1] at (3.0,3.2) {2}; 
\node[scale=1] at (3.0,2.5) {2}; 
\node[scale=1] at (2.9,4.5) {2}; 
\node[scale=1] at (2.2,4.7) {2}; 
\node[scale=1] at (4.7,4.4) {2}; 
\node[scale=1] at (3.9,4.0) {2}; 
\node[scale=1] at (3.9,2.5) {2}; 
\node[scale=1] at (4.4,3.6) {2}; 
\node[scale=1] at (2.6,4.3) {2}; 
\node[scale=1] at (4.6,4.9) {2}; 
\node[scale=1] at (4.9,2.9) {2}; 
\node[scale=1] at (5.3,2.7) {2}; 
\node[scale=1] at (3.4,3.4) {2}; 
\node[scale=1] at (4.3,1.8) {2}; 
\node[scale=1] at (3.4,1.1) {2}; 
\node[scale=1] at (5.2,1.4) {2}; 
\node[scale=1] at (2.8,2.3) {2}; 
\node[scale=1] at (3.2,2.6) {2}; 
\node[scale=1] at (4.6,1.8) {2}; 
\node[scale=1] at (5.0,0.7) {2}; 
\node[scale=1] at (2.04,1.7) {1}; 
\node[scale=1] at (3.0,1.2) {1}; 
\node[scale=1] at (3.3,0.6) {1}; 
\node[scale=1] at (4.1,-0.2){1}; 
\node[scale=1] at (0.0,3.6) {1}; 
\node[scale=1] at (0.7,3.6) {1}; 
\node[scale=1] at (1.1,4.2) {1}; 
\node[scale=1] at (1.8,5.1) {1}; 
\node[scale=1] at (5.9,2.3) {1}; 
\node[scale=1] at (4.9,0.4) {1}; 
\node[scale=1] at (4.0,0.3) {1}; 
\end{tikzpicture}
\end{center}
\end{minipage}
\caption{Student Government discussion network \cite{Hle93,dNMB04}, and the adaptation to influence graph, where edges are labeled and the label on nodes have been changed.\label{figSGD1}}
\end{figure}
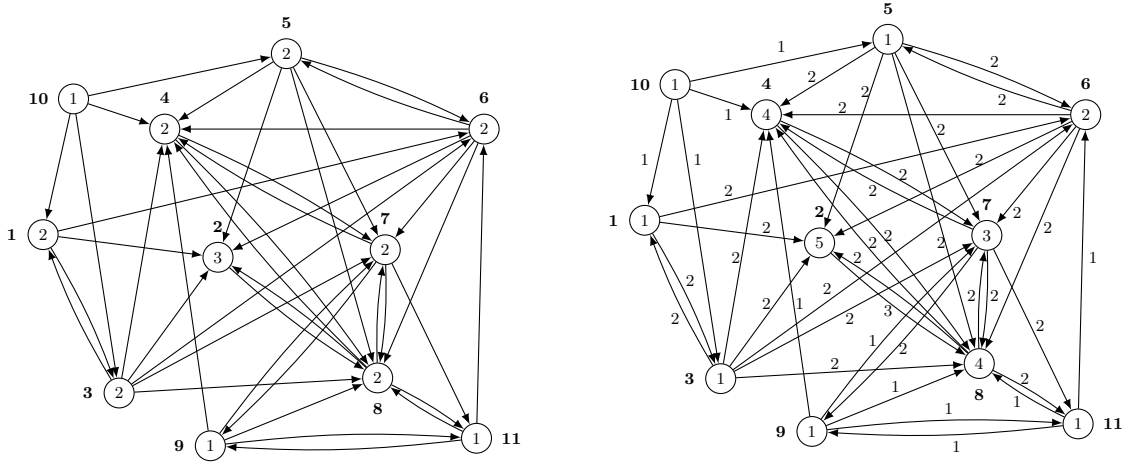

The last case of study that we analyze here starts with the social network illustrated in Figure~\ref{figSGD1}.
This network represents the communication interactions among different members of the Student Government at the University of Ljubljana in Slovenia.
Data were collected through personal interviews in 1992 and published by \cite{Hle93}, being used later by \cite{dNMB04}.

Every directed edge is a communication interaction and all of them have the same weight equal to~1. Each node is a member of the Student Government, and unlike the previous cases, here nodes are labeled beforehand: There are three {\em advisors} labeled~1, seven {\em ministers} labeled~2, and one {\em prime minister} labeled~3.

We modified slightly this network to obtain the influence graph of the second network of Figure \ref{figSGD1}. We assume that every communication interaction is an attempt to influence over another student, and the capacity to influence depends on the student's position. For instance, the advise of a prime minister does not have the same effectiveness ---marked with weight 3--- than the advise of an advisor ---marked with weight 1.
Furthermore, as the labels of the nodes should represent the difficulty of each student $i\in N$ to be influenced, according to their position in the Student Government, then they have been changed by the following values:

\begin{itemize}
  \item $f(i)=1$ 			\tabto{23ex} if $i$ is an advisor.
  \item $f(i)=\lceil deg^-(i)/2\rceil$	\tabto{23ex} if $i$ is a minister.
  \item $f(i)=deg^-(i)$ 		\tabto{23ex} if $i$ is the primer minister.
\end{itemize}

Moreover, for this network we consider a majority influence required to win, i.e., a quota $q=6$.

Table~\ref{tab:tab3} shows the results of the centrality measures corresponding to the influence game of the second network of Figure~\ref{figSGD1}.

\begin{table}[ht]
\begin{center}
\scalebox{0.9}{
\begin{tabular}{|c|c|c|c|c|c|c|c|c|}\hline
Node & $C_D^-$   & $C_D^+$   & $C_C$       & $C_B$       & {\tt Bz}    & {\tt SS}    & $C_E$ & $C_S$\\\hline
1    & 0.2       & 0.3       & {\bf 0.357} & 0.130       & {\bf 0.164} & {\bf 0.176} & {\bf 0.91} & {\bf 0.516}\\
2    & 0.5       & 0.1       & 0.200       & 0.195       & 0.154       & 0.076       & 0.45 	  & 0.515\\
3    & 0.2       & {\bf 0.6} & {\bf 0.435} & 0.169       & {\bf 0.164} & {\bf 0.176} & {\bf 0.91} & {\bf 0.516}\\
4    & {\bf 0.7} & 0.2       & 0.208       & 0.204       & 0.005       & 0.009       & 0.55 	  & 0.500\\
5    & 0.2       & {\bf 0.5} & 0.238       & 0.211       & {\bf 0.164} & {\bf 0.176} & {\bf 0.91} & {\bf 0.516}\\
6    & 0.4       & {\bf 0.5} & 0.238       & {\bf 0.304} & {\bf 0.164} & {\bf 0.176} & 0.82 	  & {\bf 0.516}\\
7    & {\bf 0.6} & {\bf 0.4} & 0.227       & {\bf 0.316} & 0.005       & 0.009       & 0.64 	  & 0.500\\
8    & {\bf 0.8} & {\bf 0.4} & 0.227       & 0.262       & 0.005       & 0.009       & 0.55       & 0.500\\
9    & 0.2       & {\bf 0.4} & 0.227       & 0.193       & 0.005       & 0.009       & 0.82       & 0.500\\
10   & 0.0       & {\bf 0.4} & {\bf 0.556} & 0.111       & {\bf 0.164} & {\bf 0.176} & {\bf 0.91} & {\bf 0.516}\\
11   & 0.3       & 0.3       & 0.227       & {\bf 0.306} & 0.005       & 0.009       & 0.82 	  & 0.500\\\hline
\end{tabular}
}
\caption{Comparison of centrality measures for the influence game version of the Student Government discussion network.
The more central values of the measures are highlighted in bold. We consider a quota $q=6$.\label{tab:tab3}}
\end{center}
\end{table}

Note that for this network, traditional measures provide different rankings.
In fact, none of the most central nodes measured with $C_C$ and $C_B$ coincide, and while the most central node for $C_C$ is the advisor~10, this is the less central according to $C_B$.
Moreover, the ministers 3 and 1 are very central for $C_C$ but with $C_B$ are at the bottom of the ranking.
This is because nodes 1, 3 and 10 have a high accessibility to all other nodes, but however, they are not good intermediaries for connecting distant nodes through paths.

Nevertheless, nodes 1, 3 and 10, as well as ministers 5 and 6, have a high score for measures {\tt Bz}, {\tt SS} and $C_S$, since the spread of the influence over the other students starting from the coalitions where they participate, is often indispensable to overcome the required quota $q$.
The same is the case for $C_E$, except for the minister 6, which is a bit less central.

Note that for any measure except $C_S$, the prime minister ---node~2--- has a relatively low centrality.
According to the measures of power indices, that is because this node has only been reported with minister~8, on which may have some influence, but he has received many interactions ---which we can understand as comments, advice, suggestions, etc.--- from other ministers and advisors, exerting a strong influence on him.
On the other hand, his low centrality with $C_E$ is explained because he can not influence any other member by himself, and at the same time his activation requires the most highest effort.
Finally, for $C_S$ this is not true because the prime minister belongs to many losing coalitions.

\section{Conclusions and future work}
\label{sec:conclusions}

Our main motivation in this work was to use influence games, which link spread of influence and decision theory, as a way to propose additional centrality measures coming from the field of cooperative game theory. We expect that such measures can be used to explain, up to some extent, the different roles with respect to social choice under diffusion process in social networks.

We have considered the framework of influence games to derive a connection between social network analysis and spread of influence in decision processes. Using this link to simple game theory we have considered four new centrality measures: {\em Banzhaf centrality measure}, {\em Shapley-Shubik centrality measure}, {\em Effort centrality measure} and {\em Satisfaction centrality measure}.
As far as we know, this is the first approach to apply power indices as centrality measures for social networks.
Our results do not contradict the relevance criteria provided by traditional centrality measures like {\em degree centrality}, {\em closeness} or {\em betweenness}. In some cases such measurements are similar to our measurements, returning expected results for a reasonable measure of centrality.
However, there are also cases where the results have been quite different from traditional measures, which may also differ significantly, as is the case of the Student Government discussion network, or the Monkeys' interaction network for the effort centrality measure.
For these cases, new reasonable centrality measures provide new approaches and insights for social network analysis.
Moreover, sometimes the new centrality measures are very eloquent for edge-labeled and vertex-labeled directed graphs, which are not supported by most measures of centrality \cite{Bor05}.

Our experiments can be extended to other power indices~\cite{Fre11}, other measures that consider power or satisfaction in societies~\cite{BRS11,BRS12}, or new measures similar or not to the {\em effort measure} or the {\em width measure}.
Moreover, it is also interesting to define other centrality measures based on the properties of influence games.

On the other hand, a similar analysis that we have introduced here can be applied to other social networks like those that exist in databases provided by \cite{Ops} or \cite{UCINET}.

Finally, there are other well known concepts related with players in simple games, such as {\em dummy}, {\em vetoer}, {\em dictators}, etc. ---see \cite{TZ99}--- that could be interesting to define as properties of actors in a social network. It seems that such concepts are closely related with their centrality on the network.
For instance, consider the advisor~10 for the analyzed Student Government discussion network of Section \ref{sec:SGd}, with a quota $q=11$ ---maximum influence required to win--- and let $X\seb N$ be a coalition. Therefore, as $deg^-(10)=0$, $\{10\}\notin X$ implies $F(X)<q$.
In cooperative game theory, this means that here the advisor~10 is a {\em vetoer} \cite{TZ99}: A coalition wins only whether it contains that significant student.

\bibliographystyle{plain}

\end{document}